\documentclass[conference,letterpaper]{IEEEtran}
\IEEEoverridecommandlockouts
\usepackage{cite}

\usepackage{amsmath,amssymb,amsfonts}
\usepackage{algorithmic}
\usepackage{graphicx}
\usepackage{textcomp}
\usepackage{xcolor}
\usepackage{array,tabularx}
\usepackage{wrapfig}
\usepackage[colorinlistoftodos]{todonotes}
\usepackage{caption}
\usepackage{subcaption}
\usepackage{hyperref}

\newenvironment{conditions*}
  {\par\vspace{\abovedisplayskip}\noindent
   \tabularx{\columnwidth}{>{$}l<{$} @{\ : } >{\raggedright\arraybackslash}X}}
  {\endtabularx\par\vspace{\belowdisplayskip}}
\def\BibTeX{{\rm B\kern-.05em{\sc i\kern-.025em b}\kern-.08em
    T\kern-.1667em\lower.7ex\hbox{E}\kern-.125emX}}
\begin{document}

\title{PSP Framework: A novel risk assessment method in compliance with ISO/SAE-21434}

\author{\IEEEauthorblockN{Franco Oberti\textsuperscript{1,2}\thanks{Authors contacts: \{franco.oberti, alessandro.savino, ernesto.sanchez, stefano.dicarlo\}@polito.it and filippo.parisi@punchtorino.com} \thanks{This work was partially supported by project SERICS (PE00000014) under the MUR National Recovery and Resilience Plan funded by the European Union - NextGenerationEU}, Ernesto Sanchez\textsuperscript{1}, Alessandro Savino\textsuperscript{1}, Filippo Parisi\textsuperscript{2}, and Stefano Di Carlo\textsuperscript{1}}
\IEEEauthorblockA{\textsuperscript{1}\textit{Control and Computer Eng. Dep., Politecnico di Torino}
Torino, Italy \\
\textsuperscript{2}\textit{PUNCH Softrontix}, Torino, Italy}
}
\maketitle

\begin{abstract}
As more cars connect to the internet and other devices, the automotive market has become a lucrative target for cyberattacks. This has made the industry more vulnerable to security threats. As a result, car manufacturers and governments are working together to reduce risks and prevent cyberattacks in the automotive sector. However, existing attack feasibility models derived from the information technology field may not always provide accurate assessments of the potential risks faced by Vehicle Electronic Control Units in different operating conditions and domains.
This paper introduces the PUNCH Softronix and Politecnico di Torino (PSP) framework to address this issue. This framework is designed to provide accurate assessments compatible with the attack feasibility models defined by the automotive product security standards. The PSP framework utilizes social sentiment analysis to evaluate the real threat risk levels.
\end{abstract}

\begin{IEEEkeywords}
Safety Critical Embedded system, Security Embedded System, Embedded System Security Threat, Threat Modeling, Road Vehicle, Natural Language Processing
\end{IEEEkeywords}

\section{Introduction}
\label{sec:Introduction}
Currently, the automotive industry is facing a number of intricate challenges. Automakers are experimenting with new technologies and advanced architectures to accommodate the new automotive frontiers driven by the green transition, with Zero-Emissions Vehicles (ZEVs) being part of the "Fit for 55" package approved by the European Community \cite{ff55}. In addition, since 2021, several ISO standards and European directives have inundated the road vehicle domain, with a focus on increasing resilience to cyberattacks. The most popular directives are UN Regulation No. 155 - Cybersecurity and Cybersecurity Management System \cite{unece-155-2021} and UN Regulation No. 156 - Software Update and Software Update Management System \cite{unece-156-2021}. These European directives require Type Approval (TA) for each vehicle or road application to gain European market access. Although certification bodies empower compliance with UNR-156 and UNR-155 for Original Equipment Manufacturers (OEMs) only, it is the OEM's responsibility to cascade the security requirements until the last company in the supply chain, thus ensuring that each vehicle component complies with European directives. To facilitate the deployment of security requirements, ISO has ratified a series of specific standards:

\begin{itemize}

\item The \emph{ISO/SAE-21434:2021 Road vehicles - Cybersecurity Engineering Standard \cite{iso21434}} introduces precise and structured security requirements for road vehicles and their components to be resilient against hacks. The standard also supports implementing the UNR155 requirements in organizations across the supply chain.

\item The \emph{ISO 24089:2023 Road vehicles - Software Update Engineering Standard \cite{iso24089}} introduces requirements and recommendations for engineering the software update process across the supply chain to support the UNR156 EU directive.

\item The \emph{ISO/PAS 5112:2022 Road vehicles - Guidelines for Auditing Cybersecurity Engineering \cite{iso5112}} provide procedures for managing audit security programs, setting audit criteria based on ISO/SAE-21434 objectives.

\end{itemize}

In January 2022, the publication of the Radio Equipment Directive (RED) \cite{red} delegated act triggered articles 3(3)(d), (e), and (f) for devices that can communicate over the internet directly or through a secondary device to improve the general level of security and the protection for personal data and privacy. With this new interpretation of radio devices, the RED impacts most of the onboard Electronic Control Units (ECU). The provision will be mandatory starting on the 1st of August 2024 through Notified Body certification that each automotive supplier shall fulfil for their devices. In 2026, the European Cyber Resilience Act (CRA) \cite{CRA} shall substitute the RED directive for Automotive supplier companies with an upcoming cybersecurity certification. Eventually, the European Parliament is reviewing the Resilience of Critical Entities (CER) Directive \cite{CRE} that proposes to expand the coverage of the Network and Information Security (NIS2) Directive \cite{nis2}, including transportation in the ten new sections. This foreshadows the automotive industry (OEMs and suppliers), seeing the automotive as an essential service provider actor, which is undoubtedly in scope for NIS2.

In this complex scenario \cite{dns,art}, this paper discusses potential inconsistencies generated using the Threat Analysis Risk Assessment (TARA) model proposed by automotive security standards during TA certification. Primarily focusing on ISO/SAE-21434:2021, the paper highlights the possible erroneous risk evaluation misled by the static model proposed by the standard and introduces preliminary ideas for making those models more tailored to evaluate the risk accurately.

The paper is organized as follows: Section \ref{sec:21434Over} overviews the ISO/SAE-21434 standard highlighting its main limitations.  Section \ref{PSP} describes the proposed solution named PSP framework from the name of the developers (PUNCH Softronix and Politecnico di Torino) and presents preliminary results obtained through a proof of concept implementation of the framework. Eventually, Section \ref{sec:conclusions} summarizes the paper's main contributions and outlines the next steps for the continuation of this work.

\section{ ISO/SAE-21434 STANDARD }
\label{sec:21434Over}

\begin{figure}[ht]
\centering
\includegraphics[width=\columnwidth]{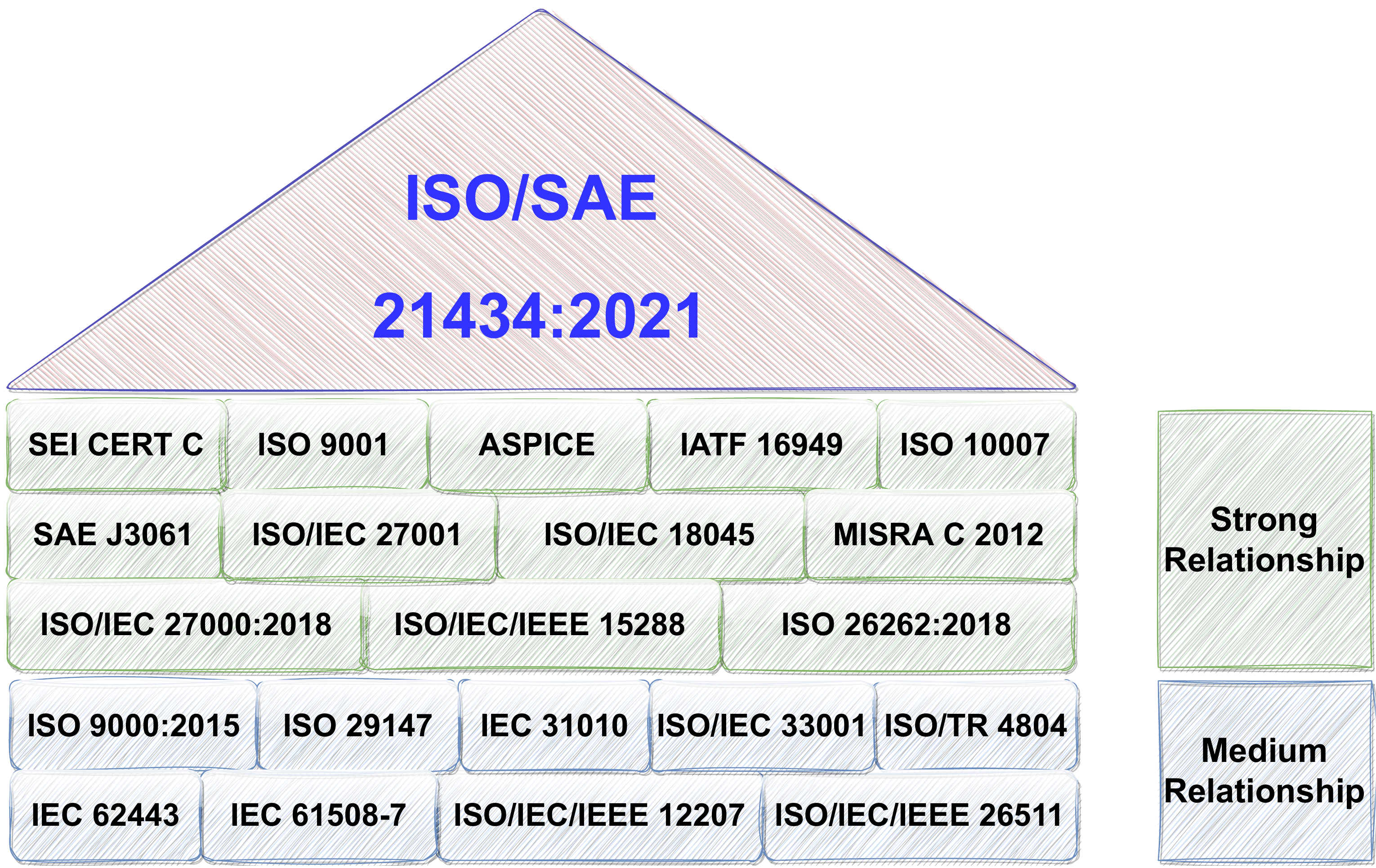}
\caption{Standards contribution list to ISO/SAE-21434}\label{ISodiag}
\end{figure}

In August 2021, the ISO/SAE-21434 was released, the first standard focused on cybersecurity for road vehicles. It was designed to support and ensure security throughout the ECU supply chain, specifically to help original equipment manufacturers (OEMs) comply with the UN R155 regulation. Figure \ref{ISodiag} displays all the standards for developing the ISO/SAE-21434 standard. It is worth noting that many of the standards used in its creation are not solely related to the automotive industry, particularly those related to cybersecurity.

\begin{figure}[ht]
\centering
\includegraphics[width=\columnwidth]{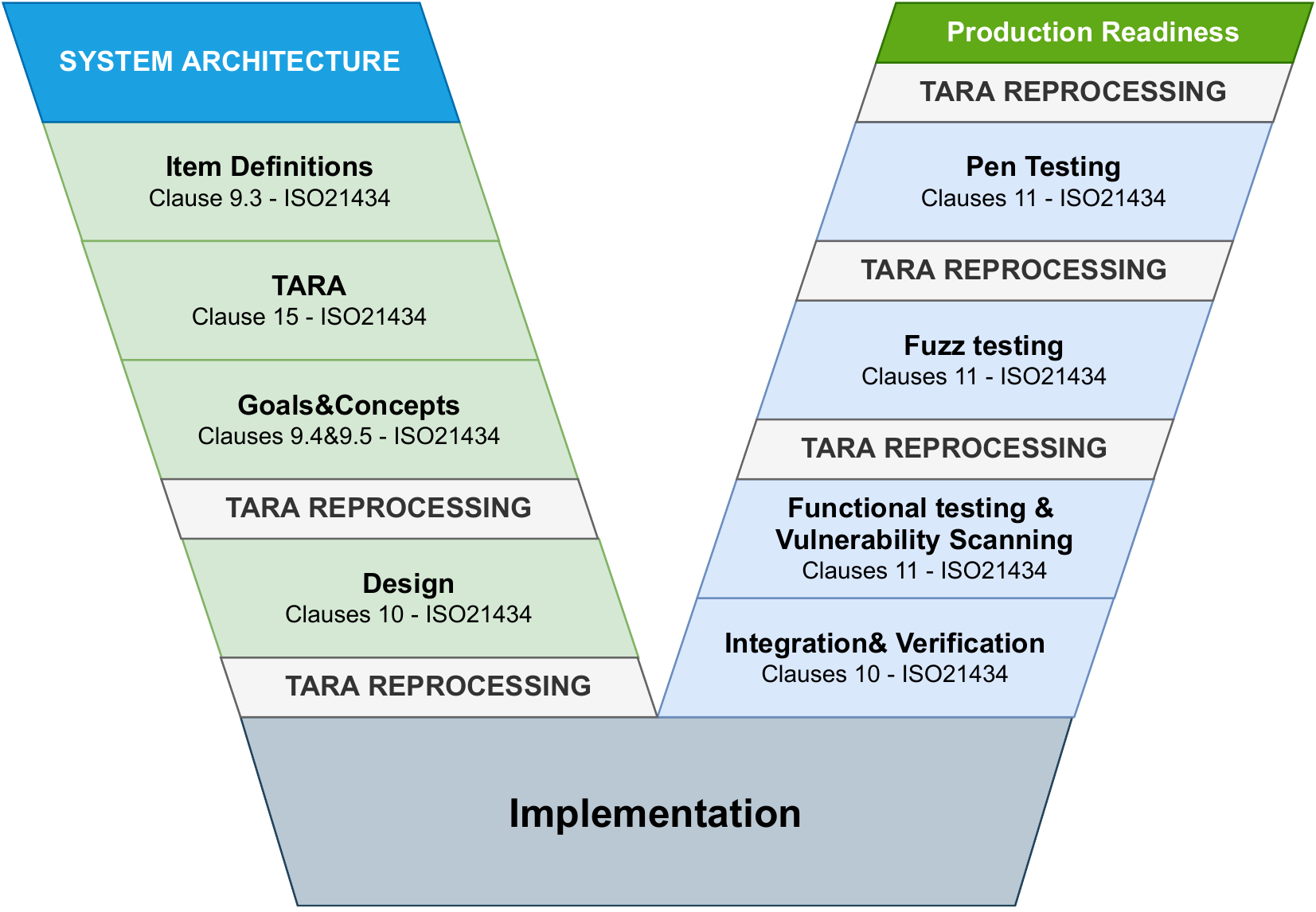}
\caption{ISO/SAE-21434 Development Life Cycle}\label{CTARA}
\end{figure}

The ISO/SAE development process is rooted in the renowned V-model widely used in software development. The ISO-26262 \cite{iso26262} standard and Automotive SPICE framework \cite{ASPICE} have also embraced it. The journey starts with creating a Threat Analysis and Risk Assessment (TARA) model, which comprises four TARA process activities: asset identification, threat scenario identification, impact rating, and attack path analysis. These activities are recursive at any point during the development cycle and are systematically applied as per Deliverable D2 in the HEAVENS project \cite{rev3}. TARA is typically called upon during production phases when a vulnerability is detected in the field. Figure \ref{CTARA} depicts the occurrence of TARA throughout the various development phases.

\begin{figure}[ht]
\centering
\includegraphics[width=\columnwidth]{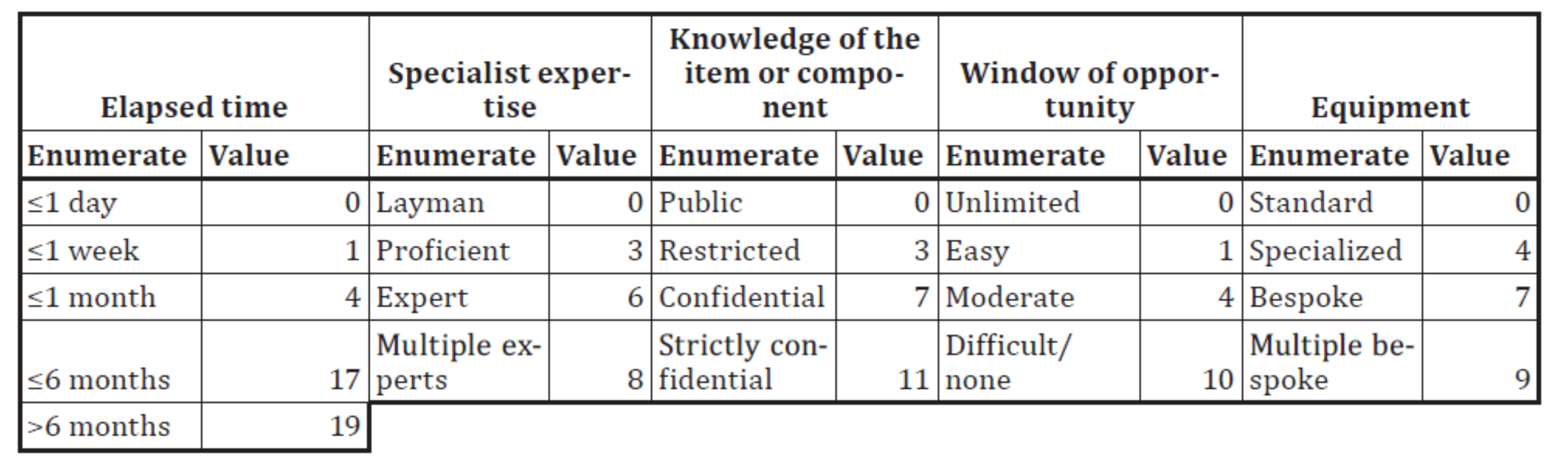}
\caption{Attack Potential weights model extracted by ISO/SAE-21434}\label{PA}
\end{figure}

 Unlike other standards like ISO-26262, the ISO/SAE-21434 provides predefined models with fixed weights defined in Clause 15 of the standard (see Figure \ref{PA}) that prevent tuning the model to fit the automotive network domain. This lack of flexibility in the model is a limitation in the TARA development that can lead to misleading results. 

\begin{figure}[ht]
\centering
\includegraphics[width=\columnwidth]{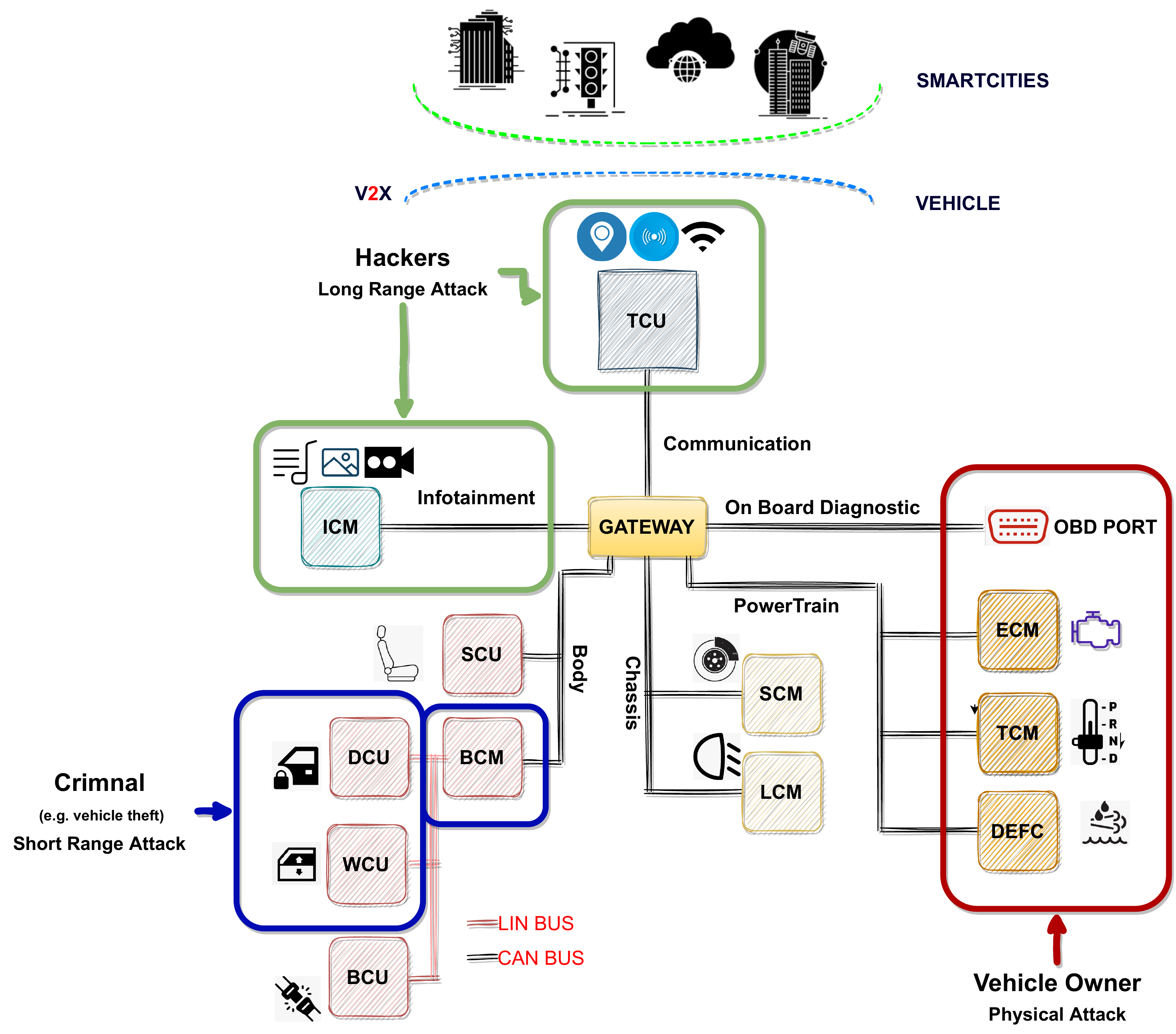}
\caption{The figure highlights in \textcolor{green}{green}  the ECUs with a suitable rate for Long-range Attack, in \textcolor{blue}{blue}  the Short-range Attack while the \textcolor{red}{red} confines the Physical Attack ECUs}\label{AttckM}
\end{figure}

The inaccurate analysis outcomes can be clarified by the influence of other security standards emphasizing IT security rather than product security. The TARA model of ISO/SAE-21434 performs well in the domains closely related to the IT Infrastructure perimeter but requires improvement in other areas. Vehicles are becoming increasingly complex, and their architecture is highly diverse, making it challenging for static modeling to function in all domains. Figure \ref{AttckM} displays a simple vehicle architecture with various functional domains and ECUs. However, not all domains are suitable for the same types of attacks. In its reports, Upstream \cite{upstream} identifies three types of attacks: long-range, short-range, and physical access. At the same time, potential attackers may have different profiles, targets, resources, and motivations \cite{Wolf2009}.

The ISO/SAE-21434 standard introduces threat feasibility models that aim to standardize threat modeling techniques across users, projects, applications, and companies. These models are significant because they promote harmonization in threat modeling. The standard defines three attack feasibility models: attack potential-based, CVSS-based, and attack vector-based. Attackers are generally classified into several categories, including Insider (such as service or maintenance personnel), Outsider (such as black hats), Rational (such as car owners), Malicious (such as criminals), Active (such as standard thieves), Passive (such as a rival/competitor), Local (such as the vehicle's owner), and so on \cite{articlepratc}.

However, using a static Enterprise IT TARA model in a complex and heterogeneous environment like the automotive industry can lead to counterproductive results. This is particularly true when developing TARA for an Engine ECU compliant with ISO-21434, as it highlights the inaccuracies in all three models when determining the attack feasibility rates.

\begin{figure}[ht]
\centering
\includegraphics[width=\columnwidth]{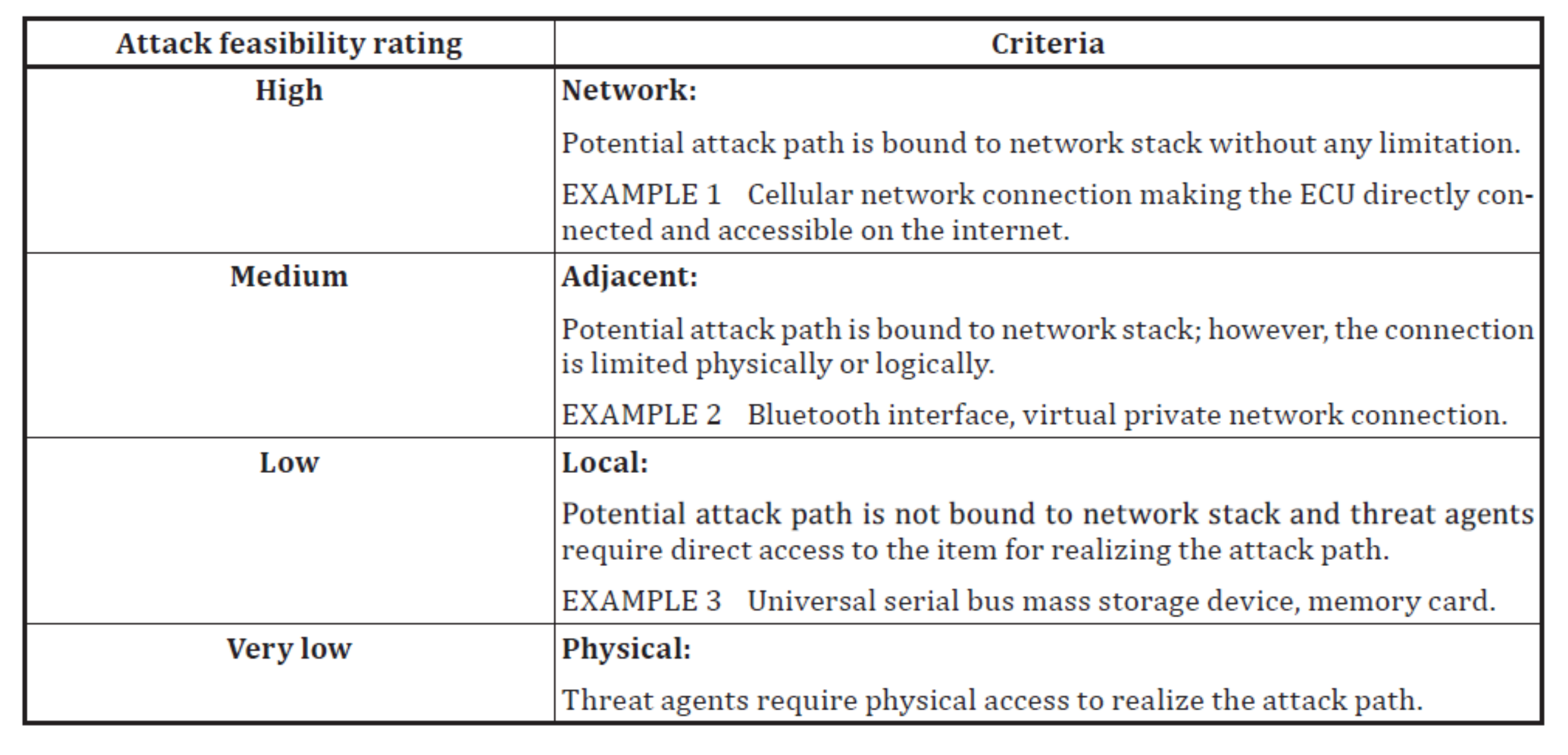}
\caption{Attack vector-based approach extracted by ISO/SAE-21434}\label{ATFR}
\end{figure}

In this particular situation, when the suggested models from the standard were applied, the result was a very high score for a remote attack and a low score for a physical attack (see an example in Figure \ref{ATFR}). However, this approach may not be suitable for automotive safety-critical hard real-time powertrain devices, where physical attacks are not rare or complex. In these systems, the primary communication occurs on the CAN bus, and external access is available through the OBD port, easily accessible in the cabin. Possible attacks on the CAN bus, particularly on the powertrain subnet domain, are physical in nature~\cite{obdpa}. Implementing a remote attack against the ECU without Firmware On The Air (FOTA) support is uncommon and challenging. Powertrain attackers usually fall into the Insider or Rational Local profile categories, which means they have unlimited time and free device access. Therefore, the ISO-21434 score produced by the ISO-21434 attack feasibility model for the powertrain scenario is misleading. Many papers report weaknesses or inaccuracies in the TARA results produced following the ISO-21434 model \cite{Escar2022,articleT}.

\begin{figure}[ht]
\centering
\includegraphics[width=\columnwidth]{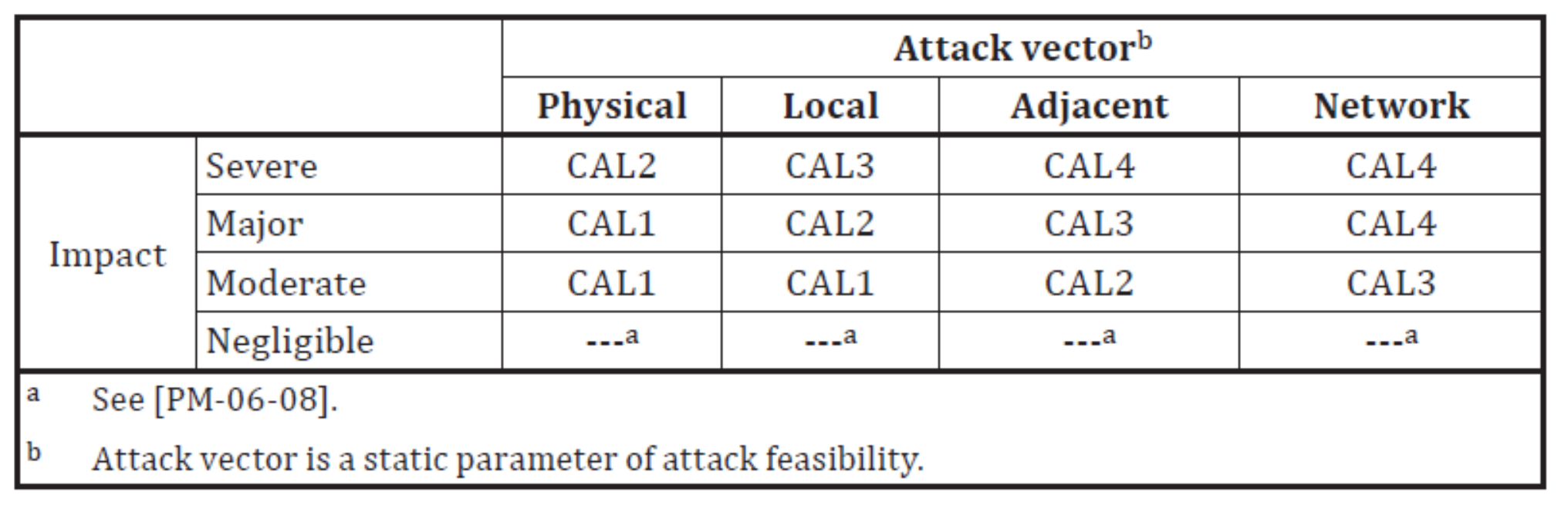}
\caption{CAL determination based on impact and attack vector parameters table extracted by ISO/SAE-21434 }\label{CALtab}
\end{figure}

Additionally, the Cybersecurity Assurance Level (CAL) is determined by attack vectors such as Physical, Local, Adjacent, and Network (Figure \ref{CALtab}). ISO-21434 defines four target security levels, with the highest being CAL4 and the lowest being CAL1. These levels correspond to the ASIL levels used in ISO-26262.

The powertrain domain oversees real-time functions that carry critical safety implications and are at risk of being targeted by Denial of Service (DoS) attacks~\cite{DoS} triggered by physical attacks. It is worth noting that such attacks do not pose a significant threat beyond CAL2 security status, which indicates a medium-low level of security emphasis. This limitation is due to the ISO-21434 standard's range of physical attack levels extending to CAL2, as illustrated in Figure \ref{CALtab}.

Therefore, many industrial technical forums require revising the TARA model applied by ISO-21434. Those papers provide solutions or improvements. However, since the system is so complex, heterogeneous, and exposed to the Man At The End (MATE) attack \cite{mate}, only some solutions can cover the entire attack surface and attacker profiles with sufficient accuracy. 

\section{PSP Dynamic TARA Model for Road Vehicle Purpose}
\label{PSP}

As mentioned in Section \ref{sec:21434Over}, the Road Vehicle sector has a lot of diversity in terms of domains, sub-domains, attack surfaces, attack vectors, and attacker profiles. Therefore, relying on fixed-weight models to evaluate security may lead to inaccurate results in certain circumstances. This situation can be detrimental to the automotive industry, as companies invest a lot of resources to make their products secure to meet legal requirements and enhance the value of their products. However, if the models used to assess security are unreliable, it can result in inefficient allocation of resources, with companies focusing on the wrong areas.
We aim to prevent such a situation by providing a non-intrusive and dynamic framework called PSP, named after its developers. This framework can assist analysts during the assessment of attack feasibility.

The PSP framework works in two distinct ways. Firstly, it utilizes active models based on ISO-21434 standard guidelines and consistently incorporates dynamic weights to evaluate all conditions. Secondly, it enables the creation of a specialized runtime model that assesses the feasibility of attacks based on financial exposure.

The PSP framework utilizes Natural Language Processing (NLP) methods to enhance the attack feasibility model. Although Machine Learning and Deep Machine Learning \cite{aigeneral} have already been implemented in various automotive applications such as Intrusion Detection Systems (IDS)~\cite{ids} and Manufacturing \cite{manufacturing}, the use of NLP in the road vehicle field is still limited \cite{escar6,NLP1}. The proposed solution uses NLP to compute sentiment classification using social media information for a subset of attacks, namely Insider, Rationale, and Local attacks described in Section \ref{sec:21434Over}. By exploiting NLP, the PSP framework can auto-generate updated weight tables that comply with ISO-21434, providing better analysis accuracy in all scenarios. This allows product security teams to provide a more precise rating on MATE attacks, which existing literature has reported as difficult to assess \cite{article}.

\begin{figure}[ht]
\centering
\includegraphics[width=0.95\columnwidth]{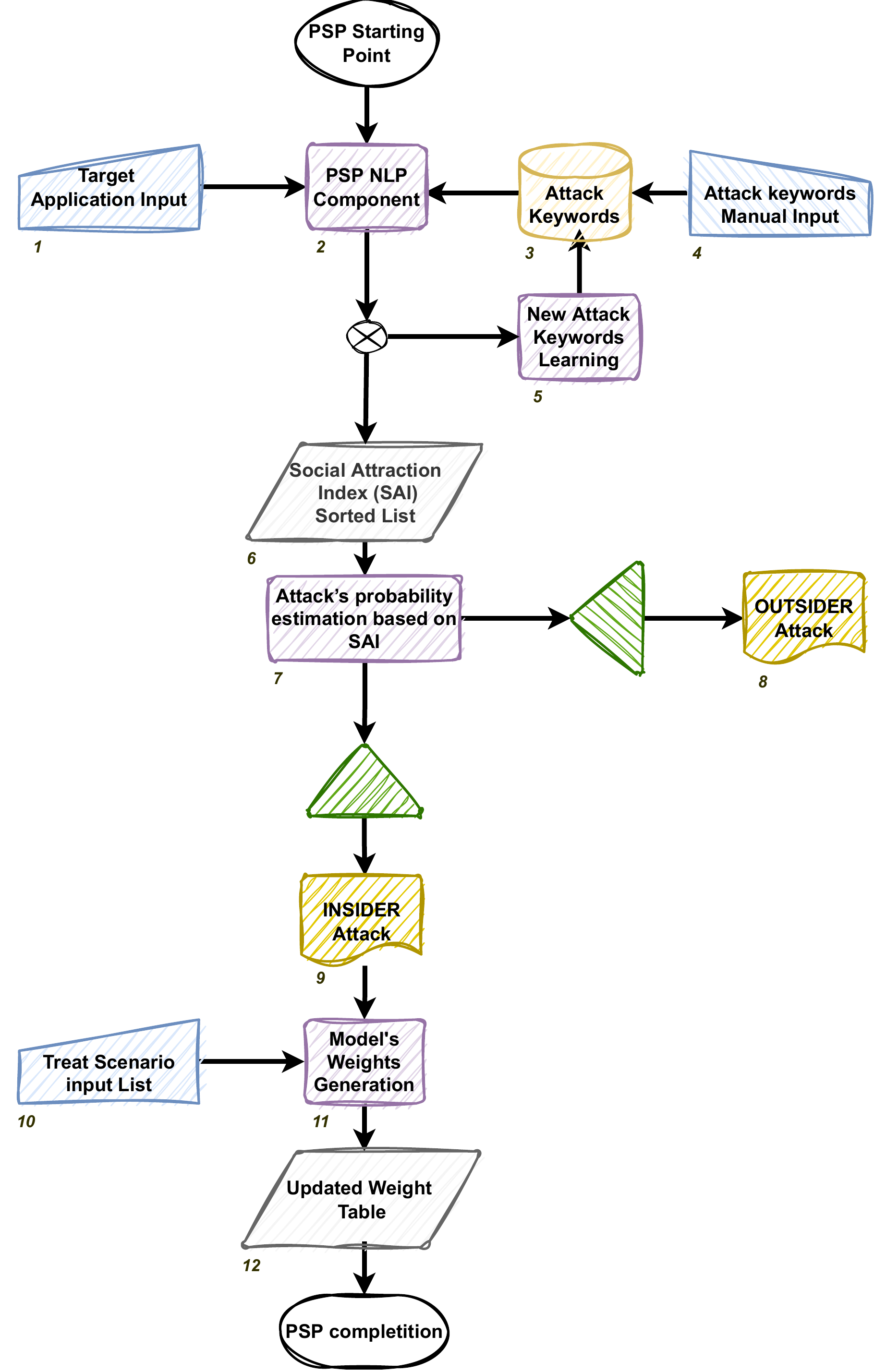}
\caption{PSP Work-Flow Scheme}\label{workflow}
\end{figure}

The diagram in Figure \ref{workflow} illustrates the PSP workflow. The initial implementation of the PSP framework uses Twitter APIs~\cite{Twitter:2023aa}. The framework takes in the target application (e.g., cars, trucks, agriculture machines), region (e.g., Europe, NA, etc.), and application category (e.g., sports cars, vans, industrial, domestic, etc.) as inputs (Figure \ref{workflow}, block 1). At the first interaction, a list of standard hashtags (e.g., \#dpfdelete, \#egrremoval, \#egrdelete, \#egroff, \#dieselpower, \#chiptuning) manually populates the keyword attack database (Figure \ref{workflow}, blocks 3 and 4).

These inputs are then processed by the PSP NLP component (Figure \ref{workflow}, block 2), which produces a sorted Social Attraction Index (SAI) list. The SAI items are calculated by querying Twitter posts based on the target application and attack keywords and elaborating on the number of views, interactions, and popularity of the identified posts. Each entry in the SAI has its attack probability estimation (Figure \ref{workflow}, blocks 6 and 7). 


While computing the SAI list, the NLP triggers a component that facilitates an auto-learning strategy to incorporate new keywords into the database for future runs. This ensures no hashtag deficiencies, which may cause partial and incomplete findings, as depicted in Figure \ref{workflow} at block 5.


Afterward, the entries in the SAI list are separated into insider or outsider categories. Here, we define insiders as all attacks that the owner is aware of and approves, even if the attack comes from third parties (e.g., an untrusted service, a racing workshop, etc.). Outsider attacks are all attacks conducted by a third party only, where the owner is oblivious (e.g., criminal attacks, thefts, black hat attacks, etc.). Typically, most threat scenarios on social media are insider, so re-tuning the standard model weight values on the outsider entries does not make sense.

The main contribution provided by the PSP framework is for all threat scenarios that belong to the insider category. These attacks are product-specific, particularly in road vehicles and IoT domains. They are new and different from the standard attacks that IT domains have experienced, requiring more experience.

At this stage, the PSP framework utilizes the insider attack list and a manually identified threat scenario list created by the product security team to generate new ISO-21434 attack feasibility tables with updated weight values (as seen in block 12 of Figure \ref{workflow}).

\begin{figure}[ht]
\centering
\includegraphics[width=\columnwidth]{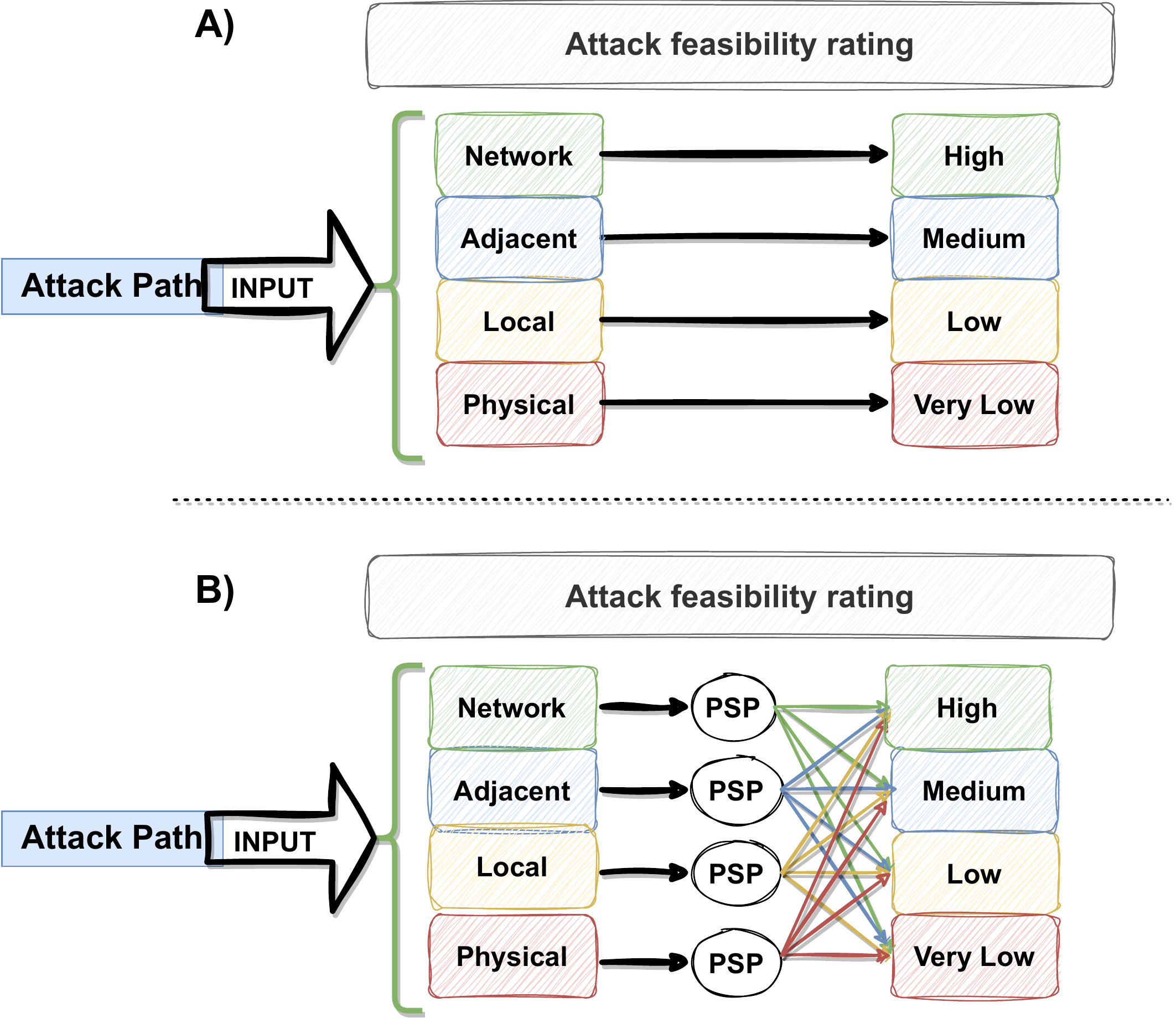}
\caption{The figure \textbf{A)} shows the attack feasibility weights, defined by ISO-21434, for outsider threats \textbf{B)}  On the contrary, the insider threats get attack feasibility weights tuned by PSP framwork} \label{atf}
\end{figure}

The ISO-21434 standard outlines constant weights for the attack vector-based approach model (as seen in Figure \ref{ATFR}). The PSP framework uses the same weight value for outsider threats, as shown in Figure \ref{atf}-A. However, for insider threats, the platform adjusts the weights by tuning them with corrective factors derived from SAI (seen in block 7 of Figure \ref{workflow}), which can change the priority of the attack vectors (as shown in Figure \ref{atf}-B).

\begin{figure}[ht]
\centering
\includegraphics[width=\columnwidth]{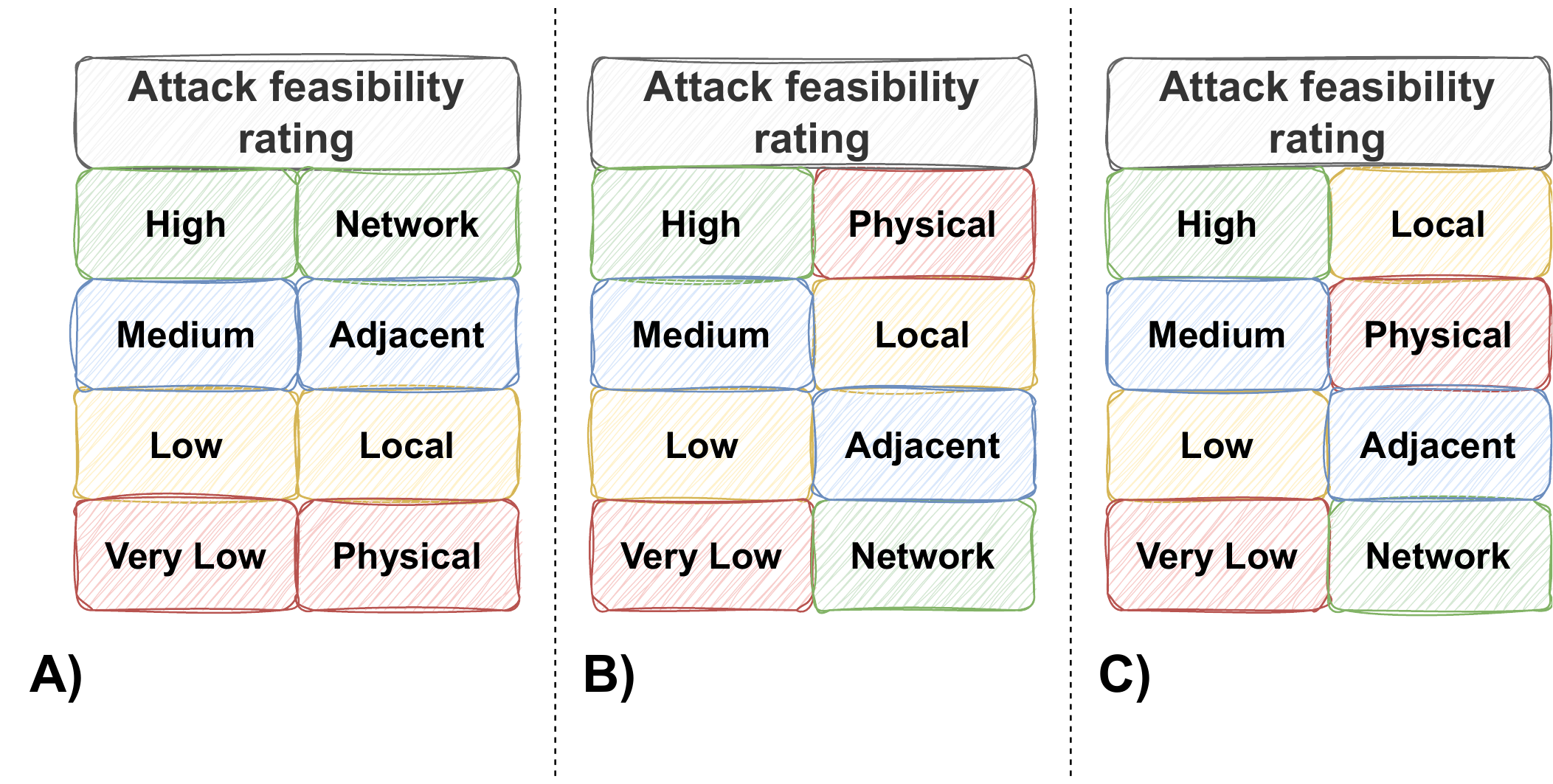}
\caption{The figure \textbf{A)} show the original G.9 table titled Attack vector-based approach provided in ISO-21434 document. Figure \textbf{B)} revised the G.9 table applying the PSP model corrections for ECM reprogramming as a Threat Scenario. The final figure, \textbf{C)}, always shows a revised G.9 table by PSP model built on the same database but limiting the data since 2022.} \label{atf2}
\end{figure}

To demonstrate the capabilities of the PSP framework, we will examine a potential threat scenario involving Engine Control Module (ECM) reprogramming. While this type of attack has a low feasibility rating according to ISO-21434 weights due to its physical attack vector (Figure \ref{ATFR}) \cite{obdphy}, according to \cite{upstream}, it has a high occurrence rate preferably based on physical attacks. By utilizing the PSP framework, we were able to update the standard attack feasibility value table, resulting in a more accurate assessment of the threat (Figure \ref{atf2}-B).

It's worth noting that the social sentiment analysis time window plays a crucial role in the PSP framework's analysis. For instance, Figures \ref{atf}-B and \ref{atf}-C show different attack feasibility ratings for the same threat scenario in the insider attack domain. This is because the PSP platform considers all Twitter posts in the former, while it only focuses on recent posts from 2021 onwards in the latter. The trend inversion highlighted by PSP began last year and is confirmed by the Upstream global automotive cybersecurity report. As a result, reprogramming via a physical attack is no longer mainstream, and attackers are more likely to opt for a local attack via OBD. While this demonstrates PSP's ability to detect current threats, it also highlights the attackers' improved techniques for bypassing secure mechanisms using local attacks.

Improving attack feasibility models, as defined by ISO-21434, is just one aspect of the PSP framework. This framework also introduces a unique strategy for creating an attack feasibility model based on a financial index. The underlying assumption is that vehicle owners initiate all internal breaches of tampering or reprogramming, even though they are illegal, to gain an advantage.

Insider attacks have established themselves as a profitable market. The significant market penetration in aftermarket tuning, such as ECU reprogramming, external control boxes, and emission defeat devices, provides access to a very lucrative market. These actions are driven either by reducing operational costs or increasing performance. Industrial vehicles fall into the first category, while standard passenger cars and light trucks belong to the second. The owner or another person must bear the cost of executing an insider attack. If the costs are reasonable for the market demand, the feasibility of that insider attack is much higher. Conversely, the attack does not match the market demand if the occurrences are infrequent.

\begin{figure}[ht]
\centering
\includegraphics[width=\columnwidth]{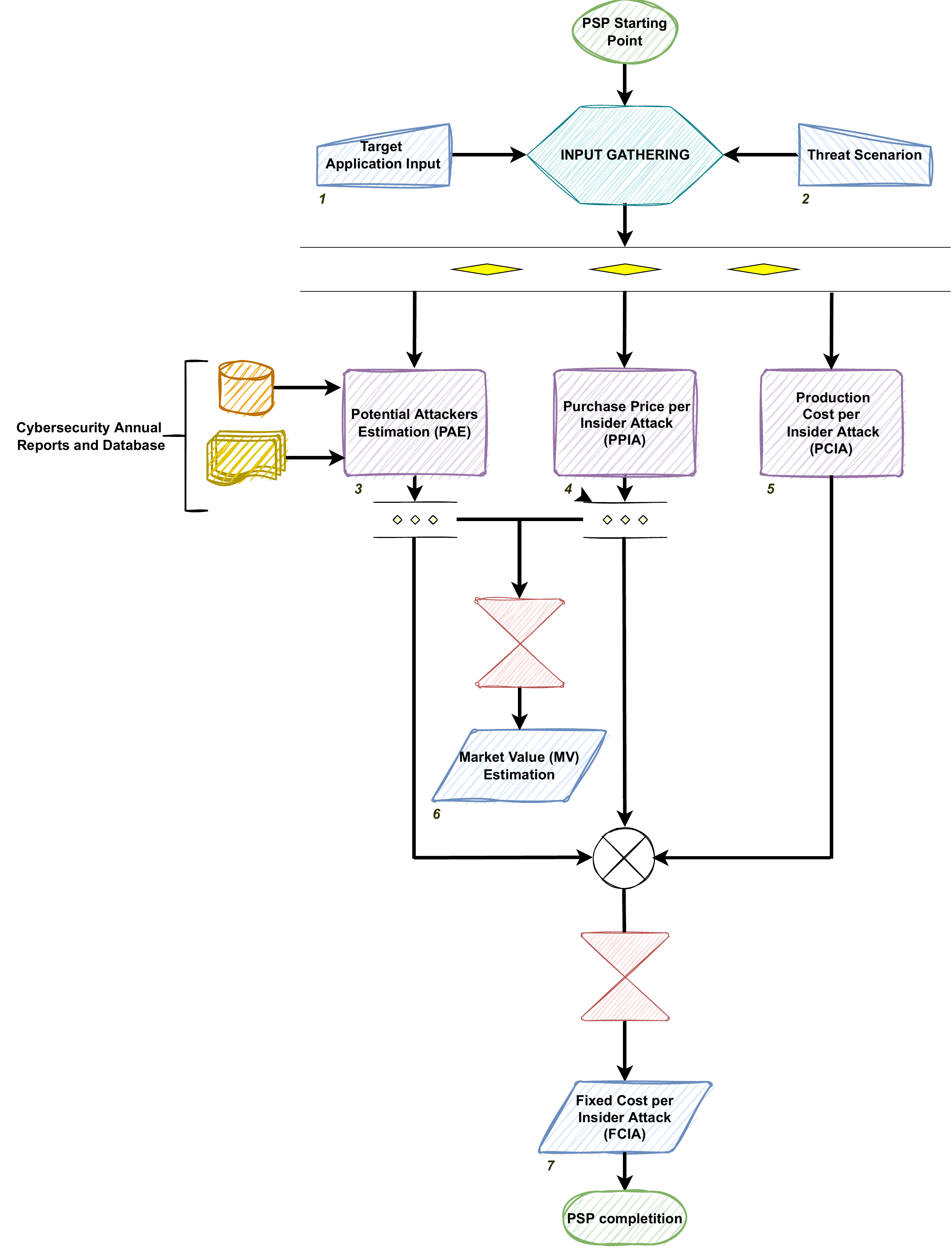}
\caption{Financial attack feasibility PSP Work-Flow Scheme }\label{workflow2}
\end{figure}

The PSP framework provides a way to map, classify, and rank all insider attacks, leveraging the high visibility offered by the market. This approach is effective, especially in assessing critical aspects with a high level of unpredictable and volatile statements led by MATE. Social tags make it easy to gather new data and instantly capture current trends quickly. Figure~\ref{workflow2} illustrates the PSP action flow.

The PSP framework is used to determine the market value ($MV$) of a potential insider attack by computing each threat scenario through equation \autoref{qq}. $MV$ is the initial measure of the size and profitability of an attack. The equation takes into account two factors: $PAE$ (Figure \ref{workflow2}, block 1) estimates the number of potential attackers, while $PPIA$ (Figure \ref{workflow2}, block 2) represents the maximum purchase price a vehicle owner would be willing to pay for an insider attack. To estimate $PPIA$, the framework utilizes NLP and text mining techniques to cluster adversary devices or services found online based on their prices. The $PAE$ value is determined by \autoref{pq}, which relies on past year's vehicle sales ($VS$) trend reports. In non-monopolistic markets, $VS$ is replaced with market share ($MS$). The framework also considers the percentage of potential attackers ($PEA$), which is determined by analyzing vehicle cybersecurity annual reports. The search parameters can be customized based on vehicle, application, years, period, historical trend, and region.

\begin{equation}\label{qq}
    MV = PAE \cdot PPIA
\end{equation}

\begin{equation}\label{pq}
    PAE  = 
   \begin{cases}        
   		VS \cdot PEA, & \text{for monopolistic markets}\\
        MS \cdot PEA, & \text{for non-monopolistic markets}
     \end{cases}
\end{equation}

\begin{figure}[ht]
\centering
\includegraphics[width=0.95\columnwidth]{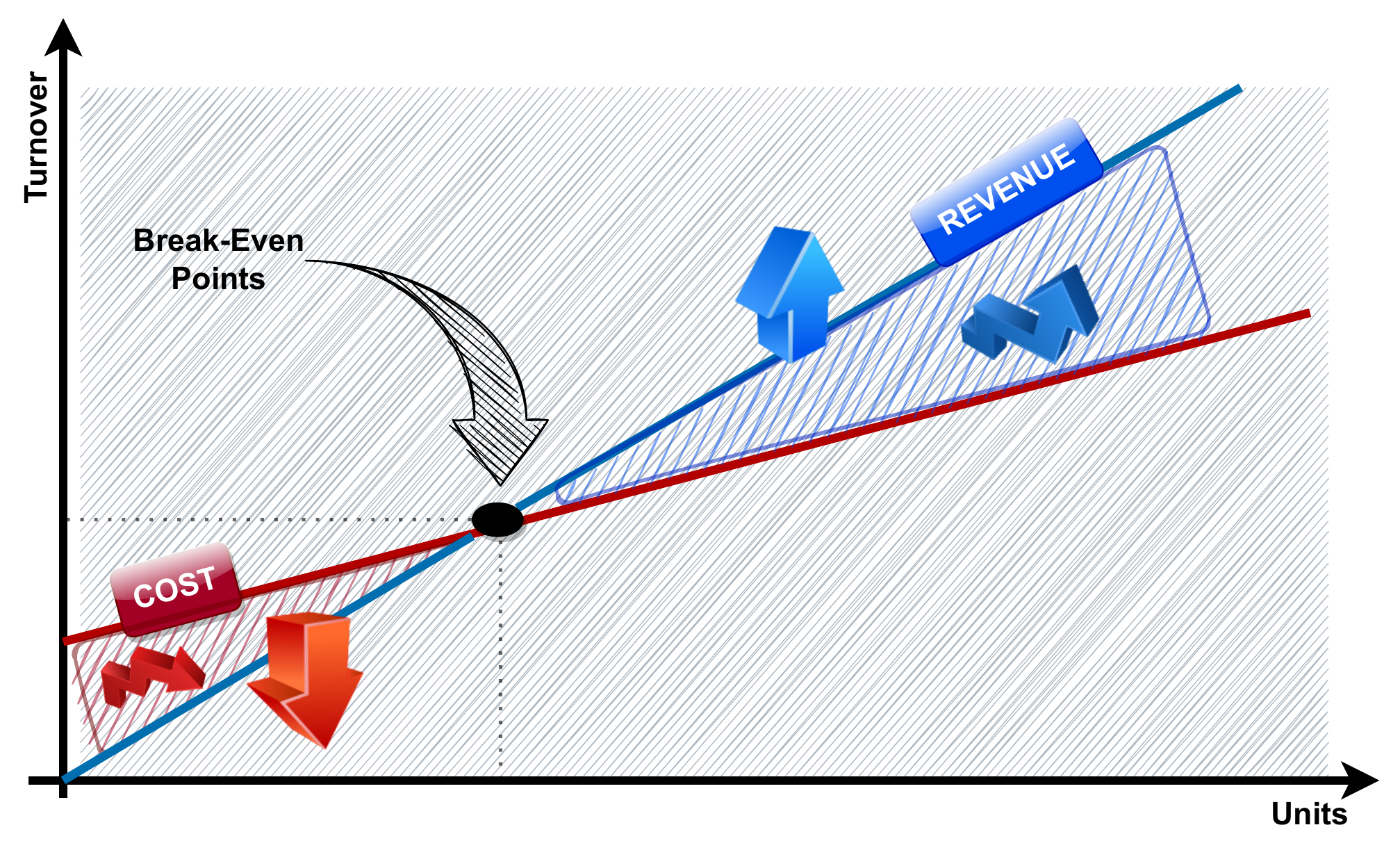}
\caption{The figure shows a standard BEP diagram. In our study, that is important to understand where the zone is profitable for the attackers.} \label{bep1}
\end{figure}

The $MV$ index helps us determine whether an attack falls within the intended scope. To increase our confidence level in estimating the feasibility of an attack, it is crucial to calculate the break-even point (BEP) using mathematical methods~\cite{BEP}. The BEP is the point at which the cost of producing an asset, in this case, an insider attack, is equal to its purchase price. Insider attacks are profitable in the blue area (shown in Figure \ref{bep1}), where their feasibility rate ranges from medium to high. Conversely, attacks in the red zone are not profitable, as their revenue is lower than their costs.
 
In \autoref{bep}, a formula is presented for calculating the $BEP$. This formula uses a numerator that represents the fixed cost ($FC$) and a denominator that represents the difference between the purchase price per unit ($PPU$) and the variable cost per unit ($VCU$). The $VCU$ considers the manufacturing cost, such as the cost of installing a defeat device in the case of an insider attack. In the scenario discussed in this paper, the $PPU$ is defined as $PPIA$ in \autoref{qq}, which is the highest price that an attacker would be willing to pay.

Since it is unlikely that a single attacker would be able to conduct a global physical attack, the revenue per unit expressed in the denominator is divided by the number of attackers ($n$), which is equivalent to multiplying the fixed cost ($FC$) by $n$.

\begin{equation}\label{bep}
    BEP=\frac{FC}{\frac{(PPIA-VCU)}{n}} = \frac{FC \cdot n}{(PPIA-VCU)}
\end{equation}

As demonstrated in \autoref{FC}, the value of $FC$ is determined by considering the total number of hours required to organize the research and development activities for the adversary ($FTEH$). The hourly cost ($ch$) is based on a standard salary for black hat hackers. The final factor involves calculating the depreciation of Capital Expenditures (CAPEX) items on a straight-line basis ($SLD$), which includes various development tools, electronic instruments, and specialized hardware and software, primarily laboratory instrumentation such as Analyzers, Tracers, Debuggers, and Oscilloscopes.

\begin{equation}\label{FC}
    FC = \left(FTEH \cdot ch \right) + SLD
\end{equation}

The information provided by the break-even point is useful for enhancing product security. The PSP framework utilizes the inverse function (\autoref{q3}) where $FC$ is an unknown term, and the $BEP$ value is equivalent to the $PAE$. In this way, the framework allows for calculating the total investment required to develop an insider attack.

\begin{equation}\label{q3}
    FC=\frac {BEP \cdot (PPIA-VCU)}{n} 
\end{equation}
\begin{figure*}[htb]
\centering
\includegraphics[width=0.9\textwidth]{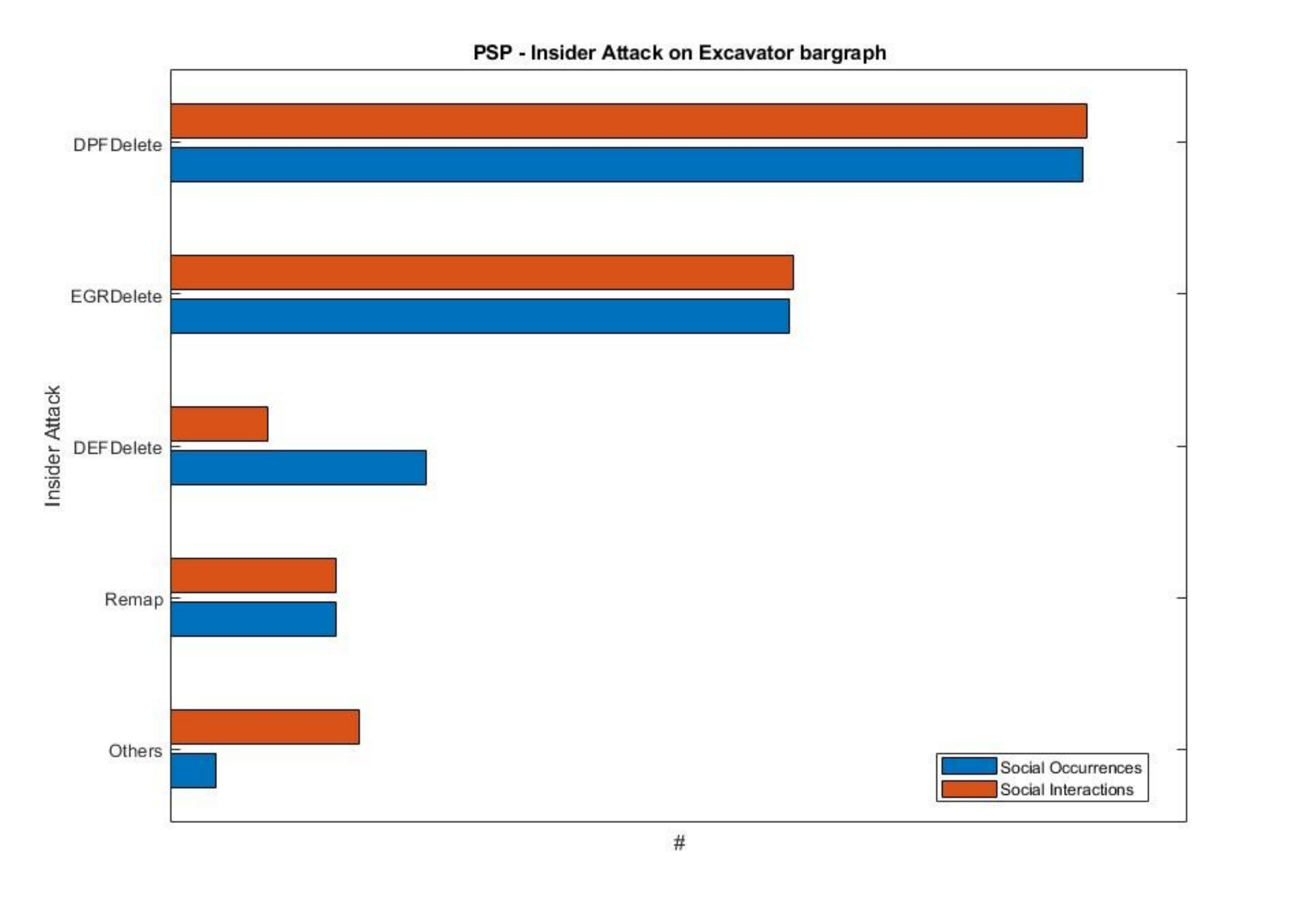}
\caption{PSP draft result about Excavator Insider Attack gotten by SAI} \label{exc}
\end{figure*}

For example, suppose we consider the query "excavator, Europe" in the PSP framework. In that case, the platform will return a picture of insider attacks concerning excavators. The SAI graph, displayed in figure \ref{exc}, reveals that disabling the Diesel Particular Filter (DPF) is the insider attack with the highest score. The PSP calculates these scores by processing and matching SAI data with the post outline, including views, occurrences, and interactions.

The estimated market value (MV) for the DPF tampering attack in Europe is 506,160 EUR per year (as referenced in \autoref{q1}). This figure is based on sales data from the previous year and refers to DPF tampering incidents on European soil excavators. The NLP method determines that a defeat device's average cost is 360 EUR after network screening. Text mining on cybersecurity reports provides the number of potential attackers, which is 1,406 for a major company based on the Upstream annual report.

\begin{equation}\label{q1}
\begin{split}
MV = PAE \cdot PPIA \\ = 1,406  \cdot 360 \;EUR \approx 506,160 \;{EUR}
\end{split}
\end{equation}

Simultaneously, the PSP platform returned an approximate value of 145,286 EUR for $FC$ as indicated in \autoref{q2}. This calculation considers the difference of 310 EUR between $PPIA$ and $VCU$, provided by the PSP platform's NLP search. It assumes three potential competitors for the attack (as estimated by the same NLP search).

\begin{equation}
\begin{split}
\label{q2}
	 FC = \frac{BEP \cdot (PPIA-VCU)}{n} \\ 
	 = \frac {1,406 \cdot 310}{3} \approx 145,286 \;\text{EUR}
\end{split}
\end{equation}

The value of $FC$ reflects the investment required for an attacker to execute an insider attack successfully. The higher the $FC$, the less feasible the attack becomes, significantly when the cost outweighs the potential revenue. In light of this example, the development team should create a secure anti-tampering DPF architecture to ensure product security that can withstand an adversary's investment of up to 145,286 EUR without being compromised.

This illustrates how the $FC$ index computed by the PSP platform can serve as a new attack feasibility index integrated into the general ISO-21434 models discussed earlier, fine-tuning market demand to better reflect the attack trend.

\section{Conclusion}
\label{sec:conclusions}
This paper presented an analysis of improving current models for assessing the feasibility of attacks in the automotive industry and introducing a new financial-based model. The main aim of the PSP platform is to move from static risk assessment models, as outlined in ISO-21434, to a runtime model environment. This approach allows for monitoring internal risks while avoiding uncertainty in all areas of the automotive sector. This flexibility ensures the models are adaptable and suitable for all vehicle domains.

The preliminary PSP framework concept presented in this paper was developed using Twitter APIs and has shown satisfactory results in terms of model quality. However, significant work must be done to operationalize the framework and automate the validation process. Additionally, enhancing features are needed to improve the automation of new keyword updates and to make attacker keyword strategies more resilient to poisoning.


Our next improvement plan is implementing a filtering strategy for messages to ensure we process only authentic posts and prevent attackers from poisoning the data. Additionally, we plan to expand the support of our framework to other social media platforms like Instagram. Our roadmap also includes a feature allowing us to access the deep web level to improve outsider attack analysis potentially.


\nocite{*}
\bibliographystyle{unsrt}
\bibliography{biblio}

\end{document}